# Home-Box based collaborative caching strategy: An asset for Content Delivery Networks

[1,2]Joachim Bruneau-Queyreix, [2]Daniel Négru, [3]Jordi Mongay Batalla

[1]VIOTECH COMMUNICATIONS SARL
13 rue Saint Honoré, Versailles, 78000 France
jbruneau@viotech.net

[2]LaBRI Lab. – University of Bordeaux
Talence, France
daniel.negru@labri.fr

[3]National Institute of Telecommunications, Warsaw, Poland
jordim@interfree.it

*Abstract* – Within the Future Internet, a new trend is foreseen with the creation of overlay networks composed of residential gateways (i.e. Home-Box), leveraging their storage and upload capacity in order to achieve scalable and cost-efficient content distribution. In this paper, we highlight an architecture of such a home-box overlay for Video On Demand (VOD) services, in cooperation with a network-aware request redirection and content caching strategy that optimize the resource usage at both network and client side to reduce the overall distribution cost. The proposed system is compared to existing solutions through comprehensive simulations. The results demonstrate the strong advantage of introducing such a network-aware and popularity-based caching strategy in terms of cost reduction for VOD services, especially for Content Delivery Networks.

## I. INTRODUCTION

The demand on network resources is growing every day driven by the needs of end users. Most of the time, this increase is not followed by the necessary upgrade of core networks capacity due to the important costs it incurs. A lot of technologies and architectures (such as Content Delivery Networks (CDNs) [1], Akamai [2], Proxy Servers and others) propose alternatives to these limitations. All of them are based on solutions with a single intermediate point, located as close as possible to the End User (EU), in order to achieve a better performance in the last mile connection. These solutions are not taking in account that when a large amount of End-Users (located under the same geographic area) simultaneously request the content, the nearest server will always be the delivering entity; hence this response results in content bottleneck [3]. Based on this, a solution that will take in account this redundant issue is needed.

The approach presented in this paper introduces the Home-Box (HB) component. The HB acts as the central element in the service distribution chain, aiming at enhancing today's home-gateways, by incorporating user-generated media processing and distribution to the existing network functionalities. This way, a given End-User is able to locate and consume services/content in an efficient way, taking in account context-related issues such as device capabilities, credentials, preferences, etc... Additionally, the HB provides means for an efficient content sharing among other End-Users and efficient content exchange mechanisms, based on multicast, unicast (DASH-based) HTTP streaming or peer-to-peer (P2P). Most of these features and functionalities rely on content, context and network information gathered at several layers by a cross-layer monitoring system. The concept has been adopted within the European project, ALICANTE, as part of the proposed architecture for Future Media Internet. In order to efficiently distribute services and content within the system, a logical interconnection of deployed Home-Boxes establishes an overlay network. Additionally, the HB is being provided with local storage capabilities allowing the latter to perform content caching and forwarding and enabling the Service Provider (PS) to push content. Thanks to this feature and along with the accurate caching strategy, the Home-Box overlay creates an assisting process to current CDNs (HB-assisted CDN). Indeed, the HB takes advantage of these distribution mechanisms and collaborative content caching functionalities to overcome some of the common issues that arise when a new highly popular VoD content is ingested into the system. This paper is dedicated to the presentation of this collaborative content caching strategy.

The rest of the paper is organized as follow. Section II provides some useful information on the background and related works on caching solutions. Section III describes the HB overlay solution with its architecture and the proposed collaborative caching strategy. Section IV presents the results obtained from the performance evaluation. We conclude in section VI.

## II. BACKGROUND AND RELATED WORK

Caching can be classified in three categories according to the location of caches in the network [3], namely: the browser cache, proxy cache and surrogate cache. Browser cache is located in the client host as part of the browser to exploit the temporal locality of the user's requests. This type of caching has the least/smallest benefit since the cache is usually quite small and there is no sharing between End-Users. Surrogate caches are located at the Web server side and are typically owned and operated by the Content Provider. The exact aim is to accelerate the server's performance. Concerning proxy caches, they are located in nodes between the server and the End-User, typically closer to End-Users than to the Content Provider servers. These nodes are owned by Network Providers or by companies operating caching services into the Internet. Proxy caching reduces service access time and permits to save bandwidth by bringing contents close to End-Users. Three types of caching strategy exist, from a research perspective, these cache types share many research challenges.

The benefits of caching and replication are numerous [4][5][6]:

• From the perspective of network infrastructures, these techniques decrease network traffic and thus minimize network congestions and improve performance.
• From the Content Providers point of view, caching and replication reduce their servers' workload and enable more efficiency in service availability, reliability and responsiveness.
• From the client point of view, caching permits to reduce significantly the service access latency for both popular contents (since they get them from nearby servers) and unpopular contents (since the contents are faster retrieved due to reduction of network congestion).

There is a number of potential problems related to Web caching and replication. For example, cache misses (when the content is not present in cache and has to be retrieved from the origin server) decreases the service access time due to the cache processing. Users might also consume stale/out-dated contents, if caches are not properly updated. Maximizing the benefits of caching and replication solutions requires a careful and intelligent design. Indeed, issues such as cache organization and cooperation, cache placement, decisions on the cachable contents, on when and where to place or replace contents and which cache will provide a requested content from a certain client, need to be solved [7] in order to have an efficient mechanism.

CDN is seen as the main managed approach for video content delivery over the Internet. Numbers of researches have focused on their optimization, especially addressing three main technical issues: replicas placement [8][9][10], content clustering [11] and client's requests redirection [12]. These works consider pure CDNs that rely on powerful server replicas with high connectivity and storage capacity. Compared to them, our proposal introduces a new equipment in the SPs' video distribution chain: the HB, capable of content caching and streaming, permitting, through the deployment of the overlay virtual layer, to overcome scalability and deployment cost issues of CDNs.

P2P systems represent another promising solution to overcome the scalability issue faced by VoD services. A certain number of researches have been performed leading to the deployment of P2P VoD systems [13]. However, if P2P systems achieve scalability while keeping service costs low, they also come with some limitations such as high peer churn, lack of control, reliability and network unfriendliness that rapidly result in network congestion, especially at the peering edges.

In our solution, HBs, stable components managed by SPs, are considered as peers. Hence this approach fully benefits from the scalability asset of the P2P model while overcoming the lack of control and reliability, peer churn, and heterogeneous caching capabilities that characterize it. Enabling a distributed edge content hosting is considered as the next step in content distribution paradigm [17]. Recent works have proposed architectures that rely on boxes deployed at the edges of the network, close to Users' terminals, for live video streaming services [18] or VoD services [19]. However in the latter, video contents are placed offline, which involves an additional delivery cost. In our proposed solution, the videos are placed during their consumptions by End-Users. In addition, an efficient spread of video copies among the HBs is also considered.

More recently, some studies related to peer-assisted CDNs have been proposed. This hybrid approach permits, on one hand, to offer scalability thanks to P2P approach and, on the other hand, to compensate the bandwidth imbalance and the churn of P2P systems by relying on the CDNs servers. The great advantages of such solution have been demonstrated analytically [20], by simulation [21], or by large-scale deployment [22]. Our proposal differs from these works by relying on the User Environment's storage and connectivity without directly involving clients' nodes participation. The HB located in the User's Environment, is the peer node acting as the always-connected "hub" between the User, Network and Service Environments, respectively being capable of shared ownership between the three actors (End-User, Service Provider, Network Provider). Therefore, the respective actors can easily manage the HB, impossible feature in case of End-User terminals. Our model is then a fully managed model and consequently overcomes the issues induced by the P2P part in such hybrid systems (e.g. free riding, reliability, peer churn issues, etc.).

III. THE HOME-BOX OVERLAY SOLUTION FOR EFFICIENT CACHING

A. Architecture

This section provides the HB entity and the HB overlay layered architecture. The HB represents a centralized element in the content delivery chain, making the bridge between the EU Environment (where the EU Terminals –EUT- are), the Service Environment (to reach all the Service Provider/Content Provider (SP/CP) sub-systems and content servers (CS)), and the Network Environment.

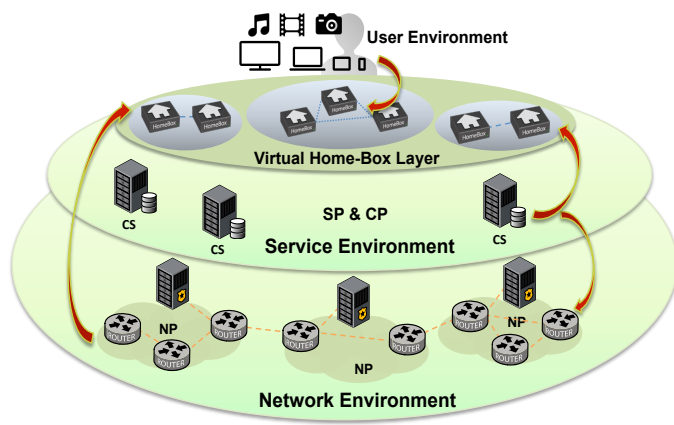

Figure 1 - Home-Box Overlay Architecture

Figure 1 briefly represents the HB overlay with the different actors composing the content delivery chain.

## B. Collaborative caching mechanism

The limitations on current distribution platform enumerated in section II are particularly relevant in the case of High-Popularity VoD use case. The former is true since the SPs are not keen to heavily invest in network infrastructures and back-office centers to distribute content that normally doesn't generate a lot of revenues. In that scenario, the simple use of a CDN presents prohibitively high costs and is not seen by the SP as the solution for this problem.

In the case of High-Popularity VoD distribution, the distributed caching at the edge nodes is seen as a very welcome add-on to the traditional cluster of VoD servers. This feature is foreseen to be a complement to the classical VOD server clusters by extending the capability of the system in terms of media distribution (would be able to handle more easily the spikes of requests).

An efficient HB caching and replication solution is therefore required in order to obtain scalable content distribution and to solve the High-Popularity VoD consumption problem. This solution is composed of (1) a collaborative caching strategy and (2) a cache management system performed by SP. Table 1 exposes the advantage of such a collaborative caching strategy compared to existing CDN solution.

| Function | CDN | HB-assisted Solution |
|---|---|---|
| Push/Pull-based Replica provisioning | Mainly push-based | Both |
| Replica provisioning algorithm | Access pattern Network topology Latency-optimized | + Popularity-based frequency + HB metrics |
| Cooperative/ uncooperative Request redirection | Cooperative | Cooperative + Adaptive delivery + Multi-source delivery |
| Cooperative/ uncooperative Caching Strategy | Uncooperative Central management Local replacement | Cooperative Collaborative online caching |

Table 1: CDN and HB-assisted CDN solution comparison

In terms of implementation, it was decided to reuse the well-known HTTP Squid proxy server, and to adapt its configuration in order to obtain the desired collaborative caching. Conceptuality speaking, the caching module is composed of the following functionalities:
- Request Redirection / peer Selection.
- Caching Management and Replacement.

The interaction with the remaining HB modules, namely at the data plane, is depicted in figure 3. The HB Caching Proxy is placed behind the HB HTTP-Client (from EU's perspective). As said before, the HB Caching Proxy is based on "Squid" implementation configured in transparent mode and enhanced with the desired caching strategy. The next configuration file sample provides an example of proxy configuration.

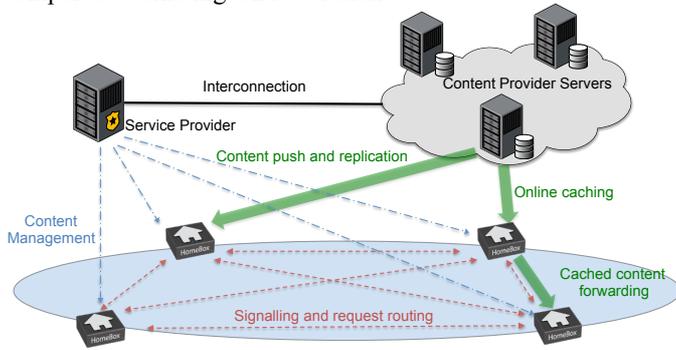

Figure 2 - Collaborative Caching Approach

Within the proposed Collaborative Caching approach, the SP may use the local HB resources, namely its storage capabilities, to push content in advance before it is being requested for consumption. The SP keeps full control of what and where this content may be pushed into. As depicted in Figure 2, The SP, through a connection to the CP servers, can push content at specific locations in the HB overlay while the HBs perform an effective collaborative caching.

The idea behind this is that if the key nodes in the network already hold a copy of the content prior to the request for consumption (le.g. the next blockbuster will be accessed tomorrow, one can predict it and push the related content to enough key nodes, based on the spike prediction), then the central VoD servers may be offloaded partially to those nodes. With time, the remaining nodes in the network will cache a copy of such VoDs on their own and will serve those copies to other nodes in the network. The next table presents a comparison of such a strategy to the traditional CDN approach.

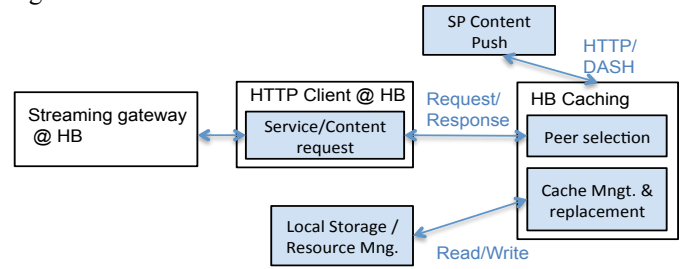

Figure 3 - HB Caching Integration

```
#set listening port and "transparent" mode
http_port 3128 tproxy
icp_port 3130

#set access control list
acl local_machines src 192.168.1.0/24
http_access allow local_machine
http_access deny all

#local cache directory config
minimum_object_size 0 KB
maximum_object_size 100 MB
cache_dir aufs /drive/squid_cache/ 2048 32 512
cache_replacement_policy lru

#set cache peer
cache_peer sibling.example.com sibling 3128 3130
proxy-only weight=X
```

The collaborative aspect of this caching strategy is taking in account some HB metrics to choose which peer to contact. The identified HB metrics are classified into two main categories: HB Resource and the HB Distance. These two categories are presented below..

HB Resource metrics: The HB Resource metrics refer to the current utilization of CPU and network resources of the HB. These metrics are monitored by a third-part module, made available via a database. Resource metrics are used by remote HBs in the End-Point/peer selection process. In this sense, candidate End-Points/peers, which have a "small" Network Distance but are currently overloaded, might not be selected if other End-Points/peers with higher Network Distance exist but in idle state. The HB resource metrics are:
- HB CPU utilization.
- Nominal max capacity of the access network interface.
- Utilization of access interface, as percentage of the max capacity.

HB Distance metrics: The HB Distance represents the metrics which define the network "cost" for the interconnection between a HB and a Content Server (or between two HBs). When a service is available from multiple SPs and/or HBs, the "nearest" HB/SP is selected (or the "nearests" when P2P is the distribution mode). The Distance Vector gathers those metrics and is defined as follow:

$$C^{(a,b)} = [c_1^{(a,b)}, c_2^{(a,b)}, c_3^{(a,b)}, c_4^{(a,b)}, c_5^{(a,b)}]$$

- $c_1^{(a,b)}$ is the number of hops traversed;
- $c_2^{(a,b)}$ is the average one-way delay (in μsec);
- $c_3^{(a,b)}$ is the average packet loss;
- $c_4^{(a,b)}$ is the average jitter (in μsec);
- $c_5^{(a,b)}$ is the percentage of duplicate packets received.

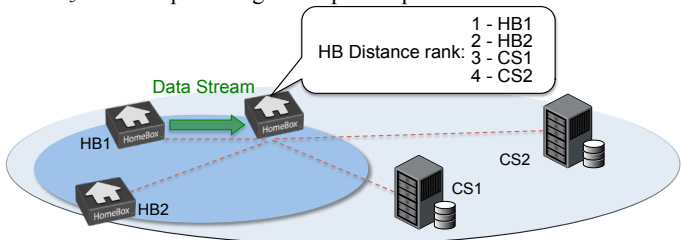

Figure 4 – Home-Box distance

Regarding the Distance Vector, it is calculated between the two edge content-aware routers, which serve the HB(s) and the SP, and does not normally include the access network. The Distance Vector depends on the IP address of the querying (client) HB, the IP address of the candidate End-Point/peer (server) SP or HB. A ranking algorithm is applied on the above-defined metrics in order to sort potential delivering HBs with respect to the requesting HB. Therefore, this mechanism supplies HBs with an ordered list of peers, each peer presenting the best service availability. Figure 4 briefly depicts this functionality.

## IV. PERFORMANCE EVALUATION

### A. Evaluation Methodology

To evaluate the proposed HB-assisted Content Delivery system, a realistic simulation environment was built based on real-world network topology, statistical studies on existing VoD services, and widely adopted system configuration. The network topology used in our simulations [23] is obtained through active traceroute probing. This router level Internet map contains 47k nodes and 119k links. This graph is representative enough to produce realistic results, because it has the same power law degree distribution, also observed by other Internet topological studies. A large set of HBs ($n=5000$) was randomly placed on this graph. The distance between two HBs is calculated in terms of router-level hop count.

To simulate the VoD service, we defined the number of available videos in the system, known as the catalog size ($c=500$). These videos have different popularities. In the simulations, we used the popularity distribution published by Netflix [24]. User requests for the video were randomly generated according to the Netflix' heavy-tailed popularity distribution (figure 5). Each HB implements the LRU (least Recently Used) caching strategy. Existing works demonstrate that this simple strategy is able to ensure at the same time good caching efficiency and system scalability.

In the simulations, we varied two major parameters of the collaborative caching system, i.e. node degree and cache size. The node degree ($d = [5, 60]$) is the number of active connections a HB should maintain with its neighboring HBs, in order to exchange cache index and forward user requests. The cache size is the storage capacity of each HB in terms of number of videos ($s = [5, 50]$). The metrics we observed are: 1) the hit ratio: the percentage of user requests fulfilled by the HB cache; 2) the hop count: the average distance between the client and his requested cache object.

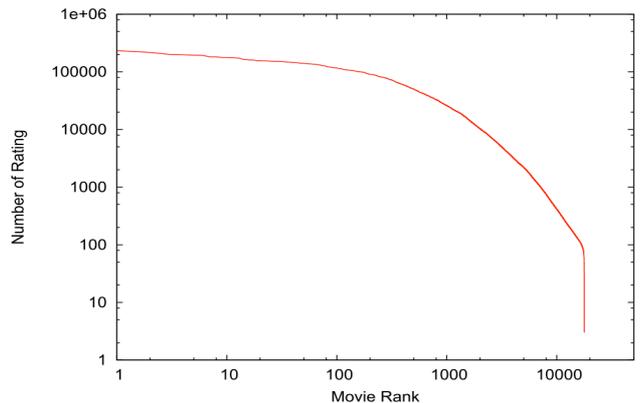

Figure 5 - Netflix Video Popularity

### B. Evaluation on caching efficiency

The caching efficiency is measured in terms of the hit ratio: when a HB receives a request from a client, it verifies firstly whether the request can be fulfilled with locally cached objects (local hit). If not, the request is forwarded to its neighbors (peering hit). If the requested video is neither cached by peering HBs (cache miss), the request will be forwarded finally to the origin server, where the entire VoD catalog is always available.

Results in figure 6 show that increasing the node degree is an efficient way to increase the hit ratio of the collaborative caches. Without the collaborative caching, the local hit ratio is only 4,5%. The hit ratio increases quickly with the number of neighbors, for example, with d=40, the total hit ratio (local+peering) is above 80%. However, due to the duplicated objects in different HBs, the hit ratio does not increase proportionally with d. On the other hand, a bigger cache size (s) will lead to both better local hit and better peering hit. But, the increase on the hit ratio is less significant compared to the previous case with node degree. We conclude that the node degree (d) has more impact on the cache efficiency that the individual cache size (s).

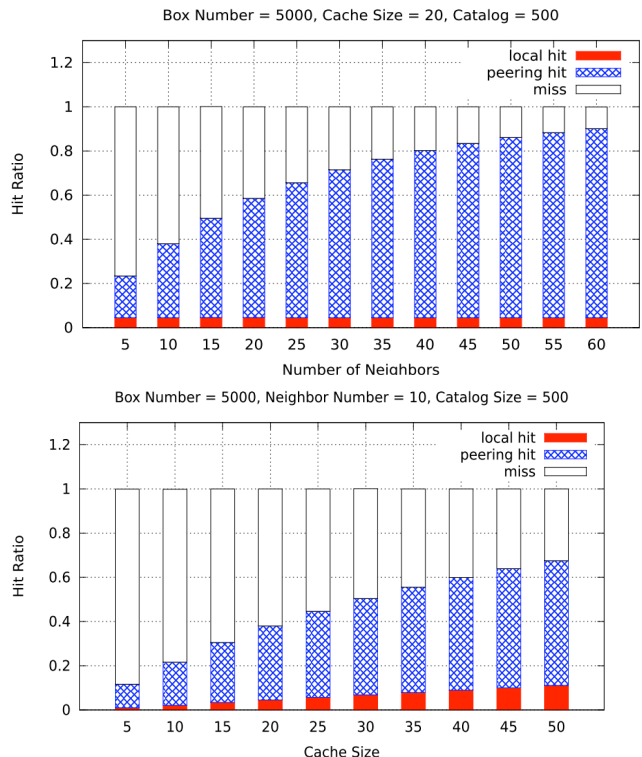

Figure 6 - Cache Efficiency

### C. Evaluation on network-awareness

Similar to other distributed systems as P2P, the proposed HB content distribution system should be "network friendly", i.e. inducing minimum unnecessary network traffic especially inter-domain traffic. To achieve this, network-aware peer selection has been implemented in our simulation, the neighbors of each HB are always the nearest ones (based on HB distance metrics, in our case hop count).Here we observe the "network friendliness" of our system, by measuring the network traffic generated by different requests.

The y1-axis of figure 7 is the average router-level hop count for each video request, which gives a good estimation of the network traffic among different HBs. The y-2 axis is the hit ratio. Both hop count and hit ratio increase along x-axis, meaning that better hit ratio will lead to more network traffic. In the first case, when a HB has more neighbors, i.e. increased reachable area, the request will be fulfilled more easily (increase hit ratio), but redirected to a further HB (increased hop count). In the second case, when a HB has more storage capacity, more requests can be fulfilled by the same set of neighbors. Therefore, the hop count increase less quickly than in the first case.

### D. Overall system efficiency

In previous section, we showed that different methods (increasing d or s) can be used to reach a better caching efficiency, but the price to pay is increased network traffic, thus less "network friendliness". In this section, we will demonstrate that there is a practical trade-off, in order to reach the optimal system efficiency.

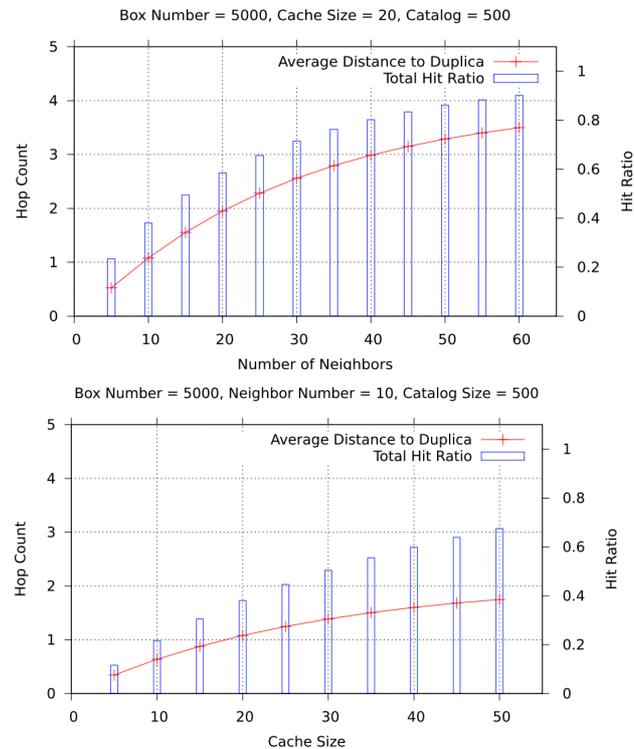

Figure 7 – Network awareness

In a VoD system with distributed caching capability, requested videos could be served by local cache, peering cache or origin server. These three cases correspond to the local hit, peering hit and cache miss in figure 6. In order to evaluate the overall system efficiency, we define a cost function $C\_overall = C\_local + C\_peering + C\_miss$, which is the sum of cost for fulfilling all requests. We consider here only network cost (i.e. hop count) for video delivery, and ignore other costs such as signaling traffic and maintenance of servers/HBs, etc. Consequently, $C\_local = 0$, and $C\_overall$ can be rewritten as a function of the total number of requests (N), the hit ratio (r) and hop count (h): . $C\_overall = N * r\_peering * h\_peering + N * r\_miss * h\_miss$. We assume a constant value of hop count for every missed request, which equals to the radius of the Internet graph used in our simulation. Thus $h\_miss = 5,5$. In this way, we can evaluate the average hop count for all requests: $C\_overall/N$, depicted in Figure 8.

The overall cost reaches its minimum at x=25. Indeed, lower node degree affects the caching efficiency, and higher node degree increases the network traffic. Section VI.C shows that the sum peering+local hit ratio with 25 neighbors is 65%

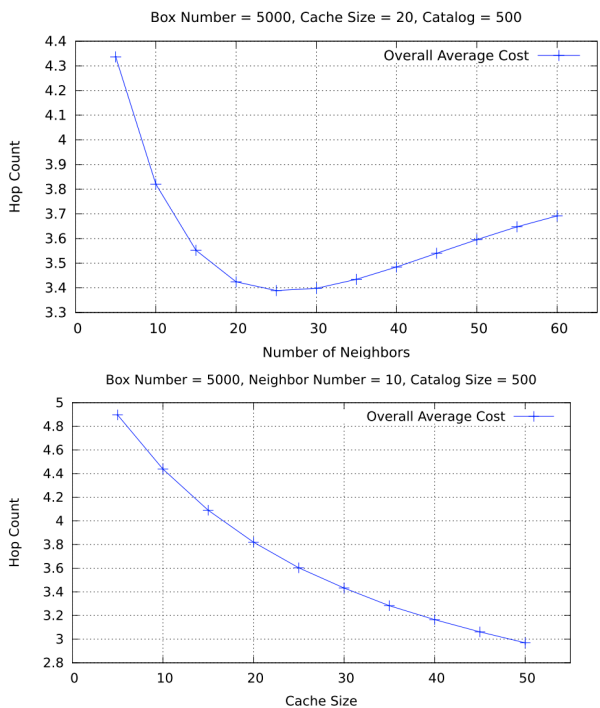

Figure 8 - Overall Cost

## V. CONCLUSIONS

The approach presented in this paper introduces the Home-Box as a constant on-line equipment with storage capability and network-awareness facilities. The HB is the central element in the service distribution chain acting as the meeting point between end-users, service/content providers and the network environments. In the context of home-gateways, the HBs provides an overlay network on top of existing content delivery network infrastructures and is part of the proposed architecture for the Future Media internet. Through an innovative popularity-based and collaborative caching strategy coupled with a proximity approach (via HB distance/resource metrics), the HB overlay off-loads the underneath network traffic by moving the latter locally. Furthermore, since the HB entity belongs to the service provider, the latter could push predicted contents in strategic locations, hence lightening the central VoD servers.

To summarize, the efficiency of proposed caching approach that consists the HB-assisted overlay has been evaluated through simulations. We observed that: 1) the hit ratio, especially the peering hit increases thanks to the collaborative caching. More requests can be fulfilled by HB overlay, and the load on origin content servers can be reduced; 2) the hop count increases with the node degree; 3) a practical trade-off can be found between the network traffic and caching efficiency. Therefore, we can conclude that the HB-assisted solution with collaborative caching can efficiently enhance the content distribution and reduce the overall cost. In the case of minimum overall cost (node degree = 25), more than 60% of requests can be fulfilled by the HB caching mechanism, instead of transiting through CDN. The virtual HB overlay is thus a good complementary solution to combine with current CDN deployment.

ACKNOWLEDGMENT

This work is part of the DISEDAN project within the European CHIST-ERA Program, which is supported by the European Union's Future & Emerging Technologies scheme (FET). We want to thank the other project partners for their support and contribution to the idea presented here.